\documentclass[epj]{webofc}
\usepackage[varg]{txfonts}   
\usepackage{hyperref}

\renewcommand*{\eqref}[1]{Eq.~(\ref{eq:#1})}

\newcommand*{\eqlab}[1]{\label{eq:#1}}
\newcommand*{\figref}[1]{Fig.~(\ref{fig:#1})}
\newcommand*{\figlab}[1]{\label{fig:#1}}

\wocname{\includegraphics[width=0.25cm,clip]{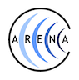} ARENA2018}
\wocname{ARENA2018}
\woctitle{ARENA2018}
\woctitle{\includegraphics[width=0.25cm,clip]{logoARENA} ARENA2018}
\begin{document}
\title{A Rotationally Symmetric Lateral Distribution Function for Radio Emission from Inclined Air Showers}
%
%

\author{\firstname{Tim} 
\lastname{Huege}\inst{1,2}\fnsep\thanks{\email{tim.huege@kit.edu}}
\and \firstname{Lukas} \lastname{Brenk}\inst{1}
\and \firstname{Felix} \lastname{Schl\"uter}\inst{1}
}


\institute{Karlsruhe Institute of Technology, Institute for Nuclear 
Physics, Karlsruhe, Germany
\and
           Vrije Universiteit Brussel, Brussels, Belgium
          }

\abstract{%
Radio detection of inclined air showers is currently receiving great attention.
To exploit the potential, a suitable event reconstruction needs to be developed.
A crucial step in this direction is the development of a model for the 
lateral distribution of the radio signals, which in the case of 
inclined air showers exhibits asymmetries due to ``early-late'' effects 
in addition to the usual asymmetries from the superposition of charge-excess and geomagnetic emission.
We present a model which corrects for all asymmetries and successfully describes the lateral distribution of the energy fluence with a rotationally symmetric function.
This gives access to the radiation energy as a measure of the energy of the cosmic-ray primary, and is also sensitive to the depth of the shower maximum.
}
\maketitle
\section{Introduction} \label{intro}

Due to the superposition of geomagnetic and charge-excess emission, the 
distribution of the energy fluence of the radio emission from extensive 
air showers is asymmetric on the ground \cite{HuegePLREP}. Lateral 
distribution functions (LDFs) have to take into account this asymmetry 
by either a two-dimensional description \cite{NellesLDF}, by 
correcting for the asymmetry induced by the charge-excess contribution
\cite{KostuninLDF}, or by treating the two contributions individually 
\cite{Glaser:2018byo}. 
In case of inclined air showers, additional asymmetries arise from 
``early-late effects'', i.e., the fact that the emission above the 
shower axis propagates longer through the atmosphere than the emission 
below the shower axis.

In this article, we first present methods to correct for the early-late 
asymmetry as well as the charge-excess-induced 
asymmetry in the energy-fluence footprints of inclined air showers.
We then propose a rotationally symmetric LDF that, when fit to the 
symmetrized energy fluences, allows precise determination of the cosmic-ray energy for inclined air 
showers.

\section{Correction of early-late effects}

Asymmetries introduced by early-late effects can be 
easily corrected to first order by assuming that the emission 
originates from a point source situated at the shower maximum, at 
geometrical distance $R_0$ from the core.
Axis distances of antennas $r_{\mathrm{raw}}$ need to be projected {\em along the line of sight from 
antenna to source} into the shower plane containing the core, resulting in the modified lateral distance $r$ as illustrated in \figref{EarlyLate}.
This is equivalent to describing the radiation pattern in terms of off-axis angles rather than axis distances.

\begin{figure}[tb]
\centering
\includegraphics[trim=2.5cm 6cm 2.5cm 5.2cm,clip,width=0.69\textwidth]{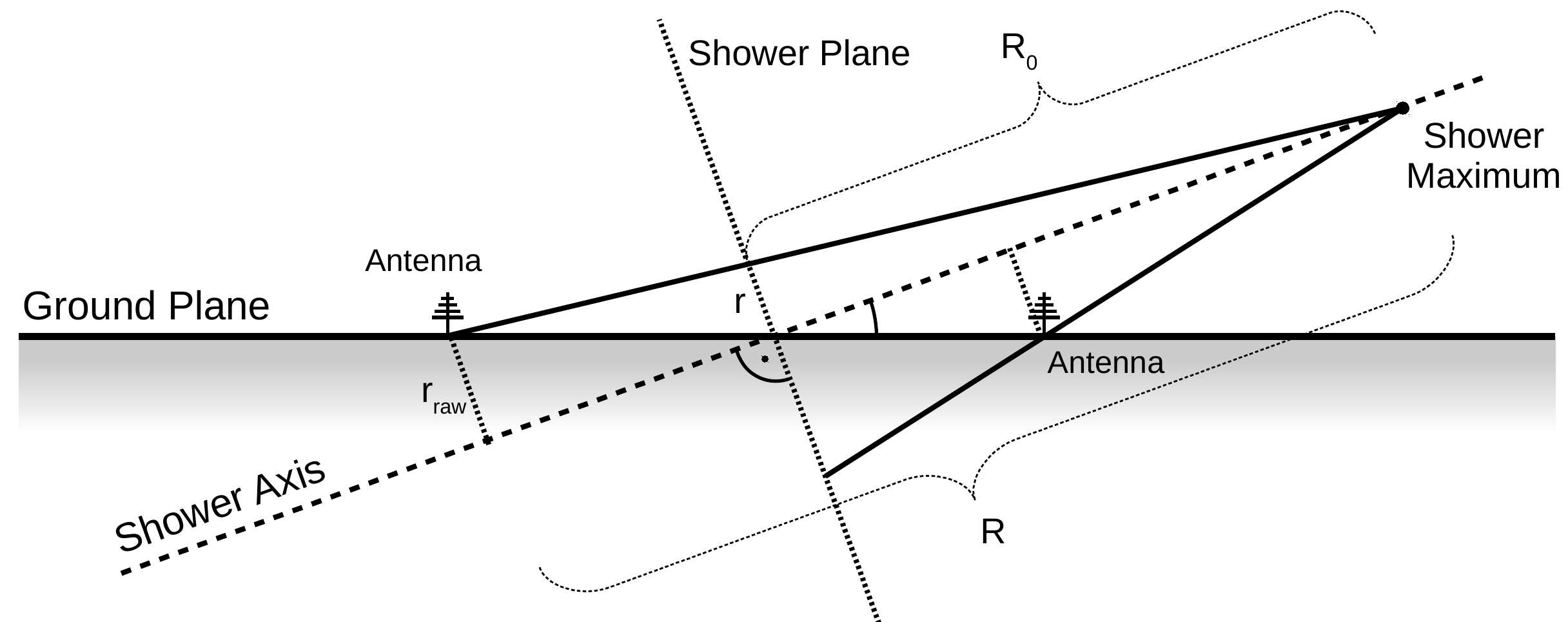}
\caption{Illustration of the correction for early-late effects.}
\vspace{-0.5cm}
\figlab{EarlyLate}
\end{figure}

The energy fluences $f_{\mathrm{raw}}$ measured in the  
antennas need to be corrected for the corresponding change of the distance 
between antenna and source using an inverse square law. (Amplitudes 
would require an inverse linear correction.) In summary:
\begin{align}
f = f_\text{raw} \cdot \left(\frac{R}{R_0}\right)^2 \quad \quad &
r = r_\text{raw} \cdot \frac{R_0}{R}
\end{align}

With the application of these two corrections, early-late asymmetries 
are effectively removed, possibly allowing the successful application 
of other approaches such as the one presented in \cite{Glaser:2018byo} to 
inclined air showers.

\section{Charge-excess correction}

After application of the early-late correction, the contribution of the 
charge-excess radiation needs to be deducted from the total energy 
fluences $f$ to determine the ``geomagnetic energy fluences'' 
$f_{\mathrm{geo}}$. This can be done in two ways.

We have derived a parameterization for the charge-excess 
fraction $a = \sin^2(\alpha) f_{\mathrm{CE}}/f_{\mathrm{geo}}$ as a 
function of zenith angle $\theta$, axis distance $r$, and depth of 
shower maximum $X_{\mathrm{max}}$:
\begin{equation}
a(\theta,r, X_\mathrm{max}) =
\cos(\theta)^{4.118}
\cdot
\left(\frac{r}{\mathrm{7937\,m}}\right)
\cdot
\exp\left(\frac{r}{1221\,\mathrm{m}}\right) 
\cdot
\exp\left(\frac{X_{\mathrm{max}}}{331.5\,\mathrm{g/cm}^2}\right).
\end{equation}

This parameterization for the frequency band from 30 to 80~MHz has been determined from CoREAS 
\cite{HuegeARENA2012a} simulations of 
air showers in the zenith-angle range from 60 to 80$^{\circ}$ spanning 
energies from 4 to 40~EeV for an 
antenna-array altitude, average atmosphere, and magnetic field configuration as valid for the site of the 
Auger Engineering Radio Array \cite{RautenbergAERA}. For other locations, the numerical parameters need 
to be re-determined.

Using the known polarization characteristics of the charge-excess and 
geomagnetic components, and assuming that both follow a rotationally 
symmetric LDF with negligible circular polarization 
\cite{GlaserRadEnergyStudy}, the geomagnetic energy fluence can then be determined 
from a measurement of the energy fluence in the polarization defined by 
the direction of the Lorentz force $\bf{v} \times \bf{B}$ as:
\begin{equation}
f_\text{geo}(r) = \frac{f_{\bf{v}\times\bf{B}}(r)}{(1 + \cos(\phi) 
\cdot \sqrt{a(\theta,r, X_{\mathrm{max}})}/\sin\alpha)^2},
\end{equation}
where $\phi$ denotes the polar angle equalling 0$^\circ$ in the 
direction given by the Lorentz force and counting counter-clockwise.

%

Alternatively, in particular if the signal-to-noise ratio of measurements in the 
polarization component perpendicular to the Lorentz force is good, the 
charge-excess fraction can be determined from the electric-field polarization of the 
signal measured at each individual antenna directly, without reverting 
to the above parameterization.

\begin{figure*}[tb]
\centering
\includegraphics[width=0.49\textwidth]{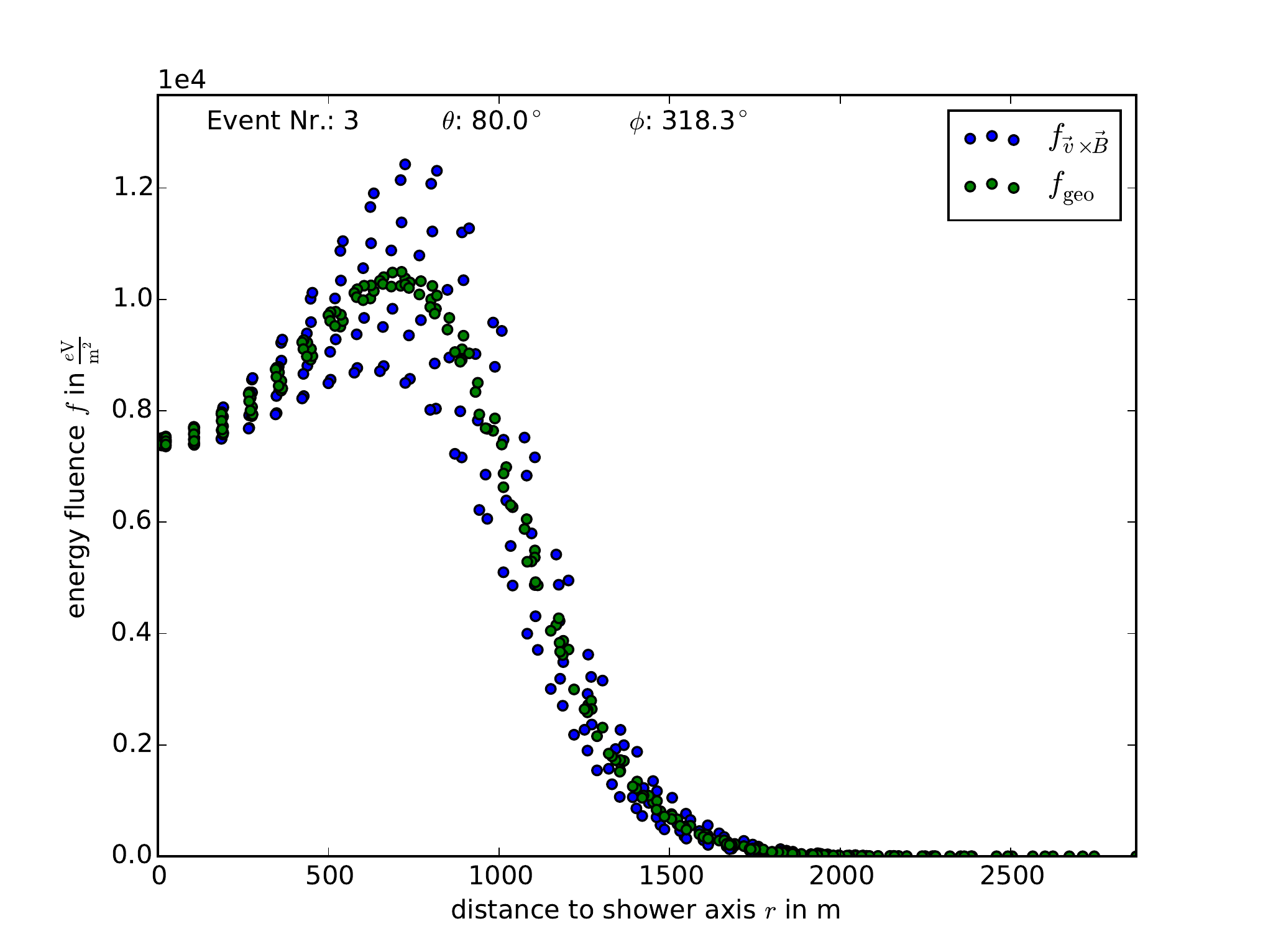}
\includegraphics[width=0.49\textwidth]{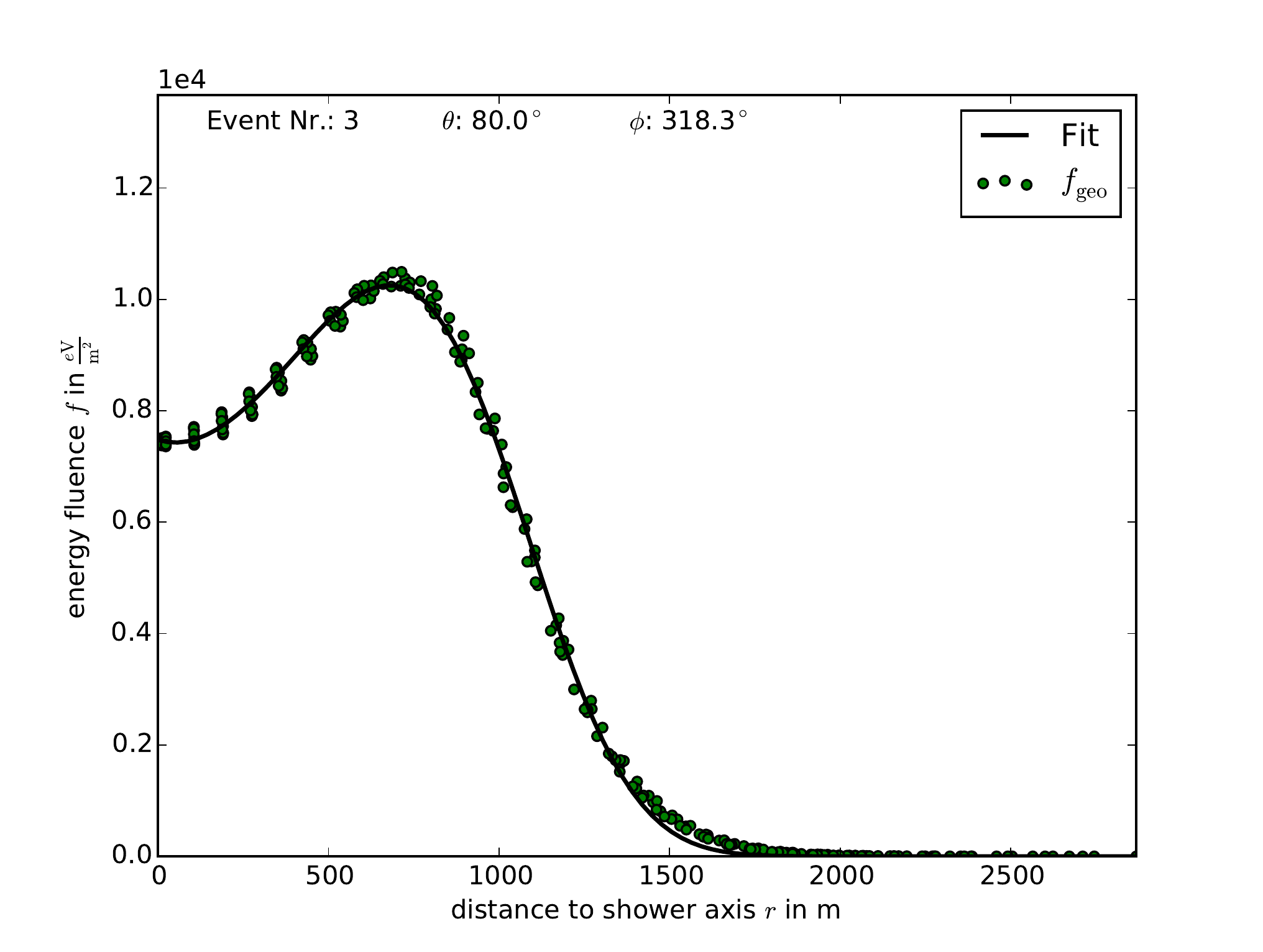}
\caption{Left: Distribution of energy fluences before (blue) and after 
(green) symmetrization through deduction of the charge-excess 
component. Right: Illustration of the LDF fit with equation 
\eqref{LDF}.}
\vspace{-0.5cm}
\figlab{LDFs}       
\end{figure*}

\section{Symmetrical lateral distribution function}

Now that the signal distributions have been symmetrized (see 
\figref{LDFs} left), a one-dimensional LDF can be readily fit to the 
energy fluences as a function of lateral distance $r$ (see \figref{LDFs} right). We use an exponential 
of a cubic polynomial as fit function. This is a generalization of the 
exponential of a quadratic function previously used by Tunka-Rex to fit measured 
amplitudes \cite{KostuninLDF}.
\begin{equation}
f_{\mathrm{geo}}^{\mathrm{fit}}(r) = A \exp\left(-Br -Cr^2 -Dr^3\right)
\eqlab{LDF}
\end{equation}

\section{Determination of the radiation energy}

After the one-dimensional fit has been applied, an area-integration can 
easily be performed to determine the radiation energy. Given that the 
charge-excess contributions to the energy fluences have been removed, 
this yields the radiation energy of the geomagnetic emission. We verified 
that, as expected, it scales quadratically with the cosmic-ray energy. 
It exhibits a spread of 10\%. Correlation with the energy in the electromagnetic 
cascade rather than the cosmic-ray energy, and application of further corrections on the air density at 
shower maximum \cite{GlaserRadEnergyStudy} are bound to further improve 
this resolution.


%
%
%
%
%

\end{document}